\documentclass[prl,twocolumn,showpacs,floatfix,superscriptaddress]{revtex4}        

\usepackage{amssymb}
\usepackage{amsmath}       
\usepackage{graphicx}
\usepackage{hyperref}
\usepackage{color}

\begin{document}

\title{\boldmath Magnetoelasticity in ACr$_2$O$_4$ spinel oxides \unboldmath}

\author{V. Kocsis}
\affiliation{Department of Physics, Budapest University of Technology and Economics and Condensed Matter Research Group of the Hungarian Academy of Sciences, 1111 Budapest, Hungary}

\author{S. Bord\'acs}
\affiliation{Department of Physics, Budapest University of Technology and Economics and Condensed Matter Research Group of the Hungarian Academy of Sciences, 1111 Budapest, Hungary}
\affiliation{Quantum-Phase Electronics Center, Department of Applied Physics, University of Tokyo, Tokyo 113-8656, Japan}

\author{D. Varjas}
\affiliation{Department of Physics, Budapest University of Technology and Economics and Condensed Matter Research Group of the Hungarian Academy of Sciences, 1111 Budapest, Hungary}

\author{K. Penc}
\affiliation{Institute for Solid State Physics and Optics, Wigner Research Centre for Physics, Hungarian Academy of Sciences, H-1525 Budapest, Hungary}

\author{A. Abouelsayed}
\affiliation{Experimentalphysik 2, Universit\"at Augsburg, D-86135 Augsburg, Germany}

\author{C. A. Kuntscher}
\affiliation{Experimentalphysik 2, Universit\"at Augsburg, D-86135 Augsburg, Germany}

\author{K. Ohgushi}
\affiliation{Institute for Solid State Physics, University of Tokyo, Kashiwa, Chiba 277-8581, Japan}

\author{Y. Tokura}
\affiliation{Quantum-Phase Electronics Center, Department of Applied Physics, University of Tokyo, Tokyo 113-8656, Japan}
\affiliation{Department of Applied Physics, University of Tokyo, Tokyo 113-8656, Japan}
\affiliation{Cross-correlated Materials Group (CMRG) and Correlated Electron Research Group (CERG), RIKEN Advanced Science Institute, Wako 351-0198, Japan}

\author{I. K\'ezsm\'arki}
\affiliation{Department of Physics, Budapest University of Technology and Economics and Condensed Matter Research Group of the Hungarian Academy of Sciences, 1111 Budapest, Hungary}

\begin{abstract}
Dynamical properties of the lattice structure was studied by optical spectroscopy in ACr$_2$O$_4$ chromium spinel oxide magnetic semiconductors over a broad temperature region of T=10-335\,K. The systematic change of the A-site ions (A=Mn, Fe, Co, Ni and Cu) showed that the occupancy of 3d orbitals on the A-site, has strong impact on the lattice dynamics. For compounds with orbital degeneracy (FeCr$_2$O$_4$, NiCr$_2$O$_4$ and CuCr$_2$O$_4$), clear splitting of infrared-active phonon modes and/or activation of silent vibrational modes have been observed upon the Jahn-Teller transition and at the onset of the subsequent long-range magnetic order. Although MnCr$_2$O$_4$ and CoCr$_2$O$_4$ show multiferroic and magnetoelectric character, no considerable magnetoelasticity was found in spinel compounds without orbital degeneracy as they closely preserve the high-temperature cubic spinel structure even in their magnetic ground state. Besides lattice vibrations, intra-atomic 3d-3d transitions of the A$^{2+}$ ions were also investigated to determine the crystal field and Racah parameters and the strength of the spin-orbit coupling.
\end{abstract}

\maketitle

\section*{Introduction}
Effect of spin ordering on the symmetry and lattice parameters of magnetic crystals, termed as magnetoelasticity, has recently attracted remarkable interest in multiferroic materials and complex magnets. This is because magnetoelasticity is closely related to the magnetocapacintace and other magnetoelectric phenomena \cite{Hemberger2005,Gnezdilov2011,Harris2006}. The dynamical coupling between lattice vibrations and spin-wave excitations plays also a crucial role in the electromagnon excitations of manganese oxides \cite{Takahashi2008,Mochizuki2010,Mochizuki2011}. For several chromium spinel oxides, the materials in the focus of the present study, multiferroic ordering has been reported at low temperatures and strong magnetic field induced changes were also found in their dielectric response \cite{Hemberger2005,Yamasaki2006,Weber2006,Mufti2010,Singh2011}. In order to reveal the nature and origin of spin-phonon coupling governing magneto-elasticity, optical probe of lattice vibrations has been shown to be an efficient tool for a broad class of chromium spinel oxides and chalcogenides with non-magnetic A-site ions \cite{Lee2000,Sushkov2005,Chung2005,Rudolf2007}.

\begin{figure}[h!]
    \includegraphics[width=7.5truecm]{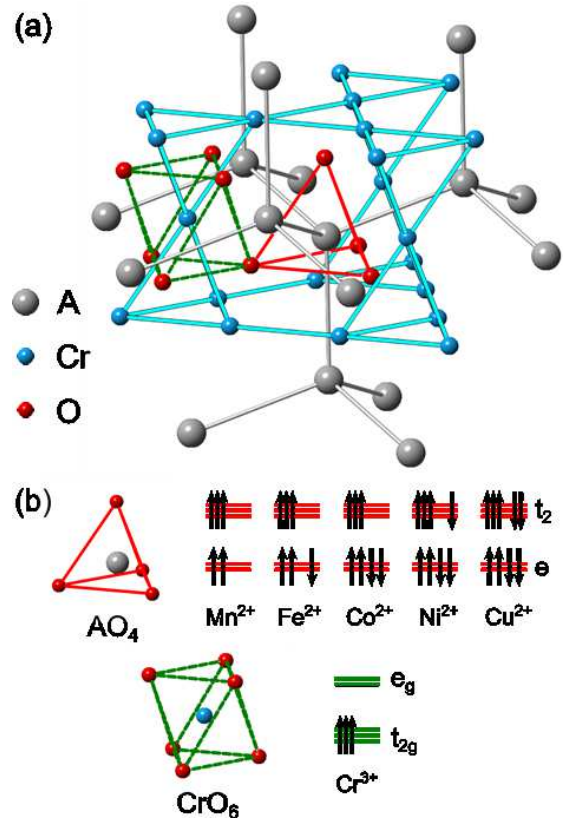}
    \caption{(a) In the structure of ACr$_2$O$_4$ spinels the Cr$^{3+}$ ions (light blue spheres) and the
    A$^{2+}$ ions (grey spheres) are located on a pyrochlore and a diamond lattice, respectively. The oxygens (red spheres) form octahedral environment
    around Cr$^{3+}$ ions, while A$^{2+}$ ions are surrounded by tetrahedral oxygen cages.
    (b) Electron configuration of Cr$^{3+}$ ion and A-site ions with increasing electron number on the 3d shell. Here, the splitting due to the local crystal field
    of the ions in the cubic phase and the effect of Hund coupling are only taken into account.}
    \label{fig01}
    \end{figure}

At high temperatures ACr$_2$O$_4$ chromium spinel oxides have the normal cubic spinel structure corresponding to the space group of Fd$\overline{3}$m as shown in Fig.~\ref{fig01}(a). Within this structure A$^{2+}$
ions form a diamond lattice with tetrahedral oxygen environment,
while magnetic Cr$^{3+}$ ions surrounded by octahedral oxygen cages
form a pyrochlore sublattice. In the local crystal field of oxygen
ions, the 3d orbitals of a Cr$^{3+}$ ion are split into a low-lying
$t_{2g}$ triplet and a higher-energy $e_g$ doublet, while the
orbitals of an A$^{2+}$ ion are split into a lower $e$ doublet and a
higher $t_2$ triplet as illustrated in Fig.~\ref{fig01}(b). The
presence of magnetic moment and orbital degeneracy on the A-site
ions can change the magnetic and structural properties in chromium
spinel oxides in two ways as compared to those members of the
ACr$_2$O$_4$ family where Cr$^{3+}$ is the only magnetic ion.

First, the antiferromagnetic J$_{Cr-Cr}$ exchange interaction
between neighboring Cr$^{3+}$ ions leads to a highly frustrated
ground state on the pyrochlore sublattice. This frustration can only
be released by the so-called spin-Jahn-Teller effect for compounds
with nonmagnetic A-site ions, where lattice distortions induce
differences between the exchange coefficients originally uniform for
each Cr$^{3+}$-Cr$^{3+}$ pairs. This removes the spin degeneracy of
the ground state and consequently long-range magnetic order can
develop as was observed at low temperatures in case of ZnCr$_2$O$_4$
\cite{Lee2000,Sushkov2005} and CdCr$_2$O$_4$ \cite{Chung2005}. In contrast,
if A-site ions are magnetic, the coupling of S=3/2 Cr$^{3+}$ spins
to the A-site spins on the bipartite diamond lattice removes the
magnetic frustration and the magnetic ordering occurs at much higher
temperatures comparable to the energy scale of the exchange
interactions.

Second, if A-site ions have orbital degeneracy in the
cubic spinel structure, it is lifted by a cooperative Jahn-Teller
distortion resulting to a tetragonal structure with space group
I4$_1$/amd already in the high-temperature paramagnetic phase
\cite{Prince1961,Tanaka1966,Crottaz1997,Kennedy2008} and the onset of
magnetic order may result in further reduction of the lattice
symmetry through a spin-lattice coupling. On the other hand, chromium oxide spinels with magnetic A
sites but without orbital degeneracy are reported to nearly maintain
the cubic structure even in their low-temperature magnetic state
\cite{Yamasaki2006,Hastings1962,Bordacs_article}. Besides spinel
chromites, the impact of orbital degeneracy on the magnetic and
structural properties is clearly manifested in vanadium oxide
spinels, where both diamond and pyrochlore sublattices can host
orbital degeneracy. The most peculiar example is FeV$_2$O$_4$
\cite{Katsufuji2008}, which exhibits successive structural
transitions due to the interplay between the Jahn-Teller effect on
the two sublattices and the influence of the spin order.

In the present paper we perform a systematic optical study of
ACr$_2$O$_4$ chromium spinels having magnetic ions also on the A
site (A=Mn, Fe, Co, Ni and Cu). In sequence of increasing number of
3d electrons on the A-site ions, A=Fe, Ni and Cu compounds are Jahn-Teller active ions, that is they show orbital degeneracy,
while MnCr$_2$O$_4$ and CoCr$_2$O$_4$ spinels have no orbital
degrees of freedom (see Fig.~\ref{fig01}(b)). Our main purpose is to
reveal the key parameters responsible for the lowering of the
lattice symmetry associated with their magnetic ordering, i.e. the
origin of magnetoelasticity in these compounds, by an infrared optical study of their lattice dynamics. We also investigate the intra-atomic \textit{3d}-\textit{3d} transitions of the A-site ions to reveal their electronic state of these magnetic ions.

Kaplan and Menyuk \cite{Kaplan2007} proposed that for J$_{A-Cr}$
(exchange coupling between spins of neighboring A$^{2+}$ and
Cr$^{3+}$ ions) sufficiently strong relative to J$_{Cr-Cr}$, the
magnetic ground state becomes unique and non-collinear magnetic
orders with three different magnetic sublattices can develop. This
has been verified experimentally for all these compounds.
CoCr$_2$O$_4$ exhibits ferrimagnetic order below T$_C$=93\,K and an
incommensurate conical spin-order develops under T$_S$=27\,K, which becomes commensurate to the lattice at
T$_{lock-in}$=13\,K \cite{Yamasaki2006,Menyuk1964,Funahasi1987,Tomiyasu2004,Ohgushi2008}. In the conical magnetic
phase the presence of spontaneous electric polarization and
ferrotoroidicity were reported \cite{Yamasaki2006}. In
MnCr$_2$O$_4$ the ferrimagnetic transition occurs at T$_C$=48\,K,
while the onset of the long-range conical order takes place at
T$_S$=14\,K \cite{Hastings1962,Tomiyasu2004,Ohgushi2008}. Similarly,
FeCr$_2$O$_4$ becomes a ferrimagnet below T$_C$=93\,K and the
conical order sets in at T$_S$=35\,K
\cite{Ohgushi2008,Shirane1964,Bacchella1964,Klemme2000}. Ferroelectric polarization was observed in this material
below T$_C$ \cite{Singh2011}. The magnetic structures of
NiCr$_2$O$_4$ and CuCr$_2$O$_4$ are rather different from the former
compounds. According to magnetization measurements NiCr$_2$O$_4$
becomes a collinear ferrimagnet at T$_C$=74\,K, where the
simultaneous lowering of the lattice symmetry to orthorhombic was
verified by X-ray scattering measurements \cite{Ishibashi2007}.
Below T$_S$=31\,K this compound exhibits a Yafet-Kittel-type canted
ferrimagnetic order \cite{Tomiyasu2004b}. In CuCr$_2$O$_4$ the
Yafet-Kittel-type magnetic order emerges right from the paramagnetic
state at T$_C$=152\,K \cite{Ohgushi2008,Prince1957}.

\section*{Experimental techniques and data analysis}

Single crystals of ACr$_2$O$_4$ for A=Mn, Fe, Co with typical
linear dimensions of 300-600\,$\mu m$ were synthesized by chemical
vapour transport, while CuCr$_2$O$_4$ was made by flux deposition
method as described elsewhere \cite{Ohgushi2008}. We adopted flux deposition method of B.~M.~Wanklyn et al.~\cite{Wanklyn1976} using solely PbO flux in order to grow NiCr$_2$O$_4$.

Unpolarized normal incidence reflectivity spectra were studied on
the (111) surface for A=Mn, Fe, Co, Ni and on the (010) surface of
CuCr$_2$O$_4$ over the temperature range of T=10-335\,K with a
Bruker IFS 66v/s Fourier-transform IR spectrometer coupled to a
microscope (Bruker Hyperion). The lattice vibrations were found to
be sensitive to polishing induced mechanical strain, thus,
reflectivity spectra were measured on as-grown surfaces at low
energies, $\omega$=100-7700\,cm$^{-1}$. The surface roughness of
MnCr$_2$O$_4$ and NiCr$_2$O$_4$ samples demanded to use of polished
surfaces in order to avoid light scattering at higher energies.
Correspondingly, reflectivity spectra were also recorded over
$\omega$=1700-24\,000\,cm$^{-1}$ on polished surfaces. Reflectivity
spectrum of each compound was measured up to
$\omega$=250\,000\,cm$^{-1}$ at room temperature
\cite{Ohgushi2008, Ohgushi_talk}.

Optical conductivity spectra were obtained from the reflectivity
data by Kramers-Kronig transformation. The low-energy part of the
measured reflectivity spectra was extrapolated to zero photon energy
as a constant value, while the high-energy part was assumed to
follow the free electron model above $\omega$=1\,000\,000\,cm$^{-1}$. The advantage of
the optical conductivity spectra versus the reflectivity spectra is
that phonon excitations appear as sharp peaks in the former, while
close resonances are more difficult to resolve in the reflectivity
data. In the analysis of the optical phonon modes we fitted the
corresponding low-energy part of the optical conductivity spectra
with a sum of Lorentzian peaks according to
    \begin{equation}
    \displaystyle \sigma \left( \omega \right)= -i\omega\epsilon_o \left( \varepsilon_{\infty} -1 + \sum_{i=1} \frac{S_i}{\omega_i^2 - \omega^2 - i \gamma_i \omega}
    \right),
    \label{eq:oscmodel}
    \end{equation}
where $S_i$, $\omega_i$ and $\gamma_i$ stand for the oscillator
strength, the resonance frequency and the damping rate of the modes,
$\varepsilon_o$ is the vacuum permittivity and $\varepsilon_{\infty}$ is the dielectric constant at high
energies.

    \begin{figure*}[t!]
    \includegraphics[height=17.truecm]{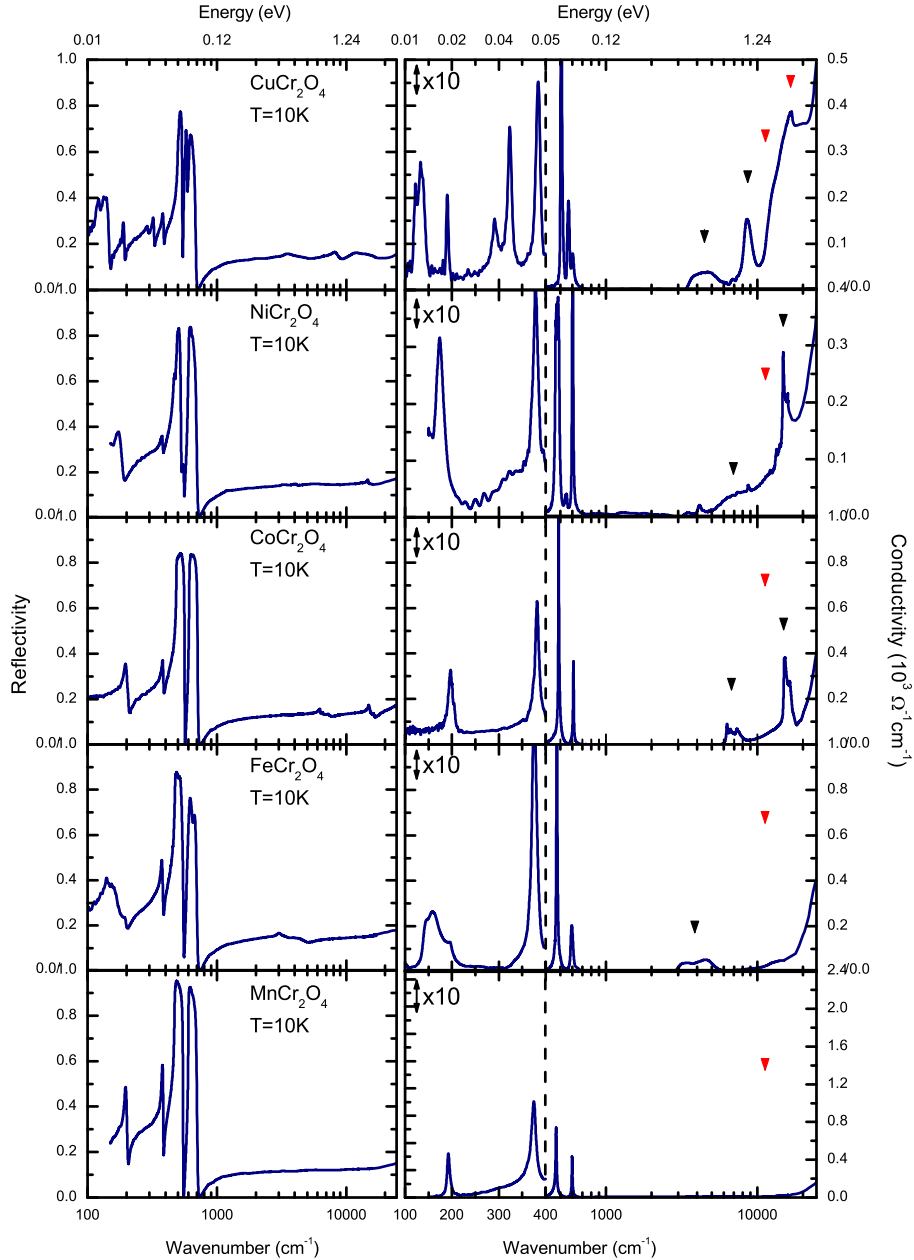}
    \caption{Reflectivity spectra (left panel) and the real part of the optical conductivity
            spectra (right panel) over the photon energy range of 100\,cm$^{-1}$ and 24\,500\,cm$^{-1}$ for the ACr$_2$O$_4$ (A=Mn, Fe, Co, Ni and Cu)
            spinel crystals measured at T=10\,K. This spectral range covers the in-gap excitations, namely the optical phonon modes and the intraatomic 3d-3d transitions.
            The 3d-3d transitions of A$^{2+}$ ions are indicated by black triangles, while the 3d-3d transitions of Cr$^{3+}$ ions are shown with grey triangles.
            The low-energy part of the conductivity spectra is ten times magnified for better visibility.
            In MnCr$_2$O$_4$ and CoCr$_2$O$_4$ only the four $T_{1u}\mathrm{(IR)}$ infrared active phonon modes corresponding to the cubic symmetry are visible. In the other compounds,
            some of these four modes are clearly split indicating the lowering of the crystal symmetry.}
    \label{fig02}
    \end{figure*}

\section*{Results}

The measured reflectivity and the corresponding optical conductivity
spectra are plotted over common energy scales for ACr$_2$O$_4$
spinels (A=Mn, Fe, Co, Ni and Cu) in Fig.~\ref{fig02} at T=10\,K. At
the high-energy side the upturn of the optical conductivity
corresponds to a band gap of $\Delta \approx$2.5\,eV characteristic to these chromium spinel
oxides. The gap value is not significantly influenced by the
variation of the A-site ions, therefore, the lowest-energy interband
transition is mainly related to the O 2p $\rightarrow$ Cr 3d charge
transfer excitations. Below the optical gap weak excitations exist
in the visible and near infrared region. Since these structures are
sensitive to the change of the A-site ion they are assigned as
intra-atomic \textit{3d}-\textit{3d} transitions of the
tetrahedrally coordinated A cations in agreement with former optical
studies \cite{Ohgushi2008}. A broad and low-intensity hump at
E=1.6\,eV is the only structure common in each spectra. This feature
is due to the \textit{3d}-\textit{3d} transitions of the Cr$^{3+}$
ions. Finally, in the low-energy range ($\omega$=100-700\,cm$^{-1}$)
optical phonon excitations characteristic to the lattice structure
are present. In the following sections dependence of the phonon excitations and the
\textit{3d}-\textit{3d} transitions on the A-site cations and the temperature are studied.

\subsection*{Lattice vibrations in ACr$_2$O$_4$ spinels without orbital degree of freedom}

According to powder X-ray diffraction measurements the lattice
symmetry for A=Mn, Fe, Co corresponds to the cubic space group $Fd
\overline{3}m$ at room temperature, while in the case of A=Ni, Cu
the symmetry is already tetragonal with the space group $I4_1 /amd$
\cite{Crottaz1997,Ohgushi2008}.

    \begin{figure}[h!]
    \includegraphics[width=8.5truecm]{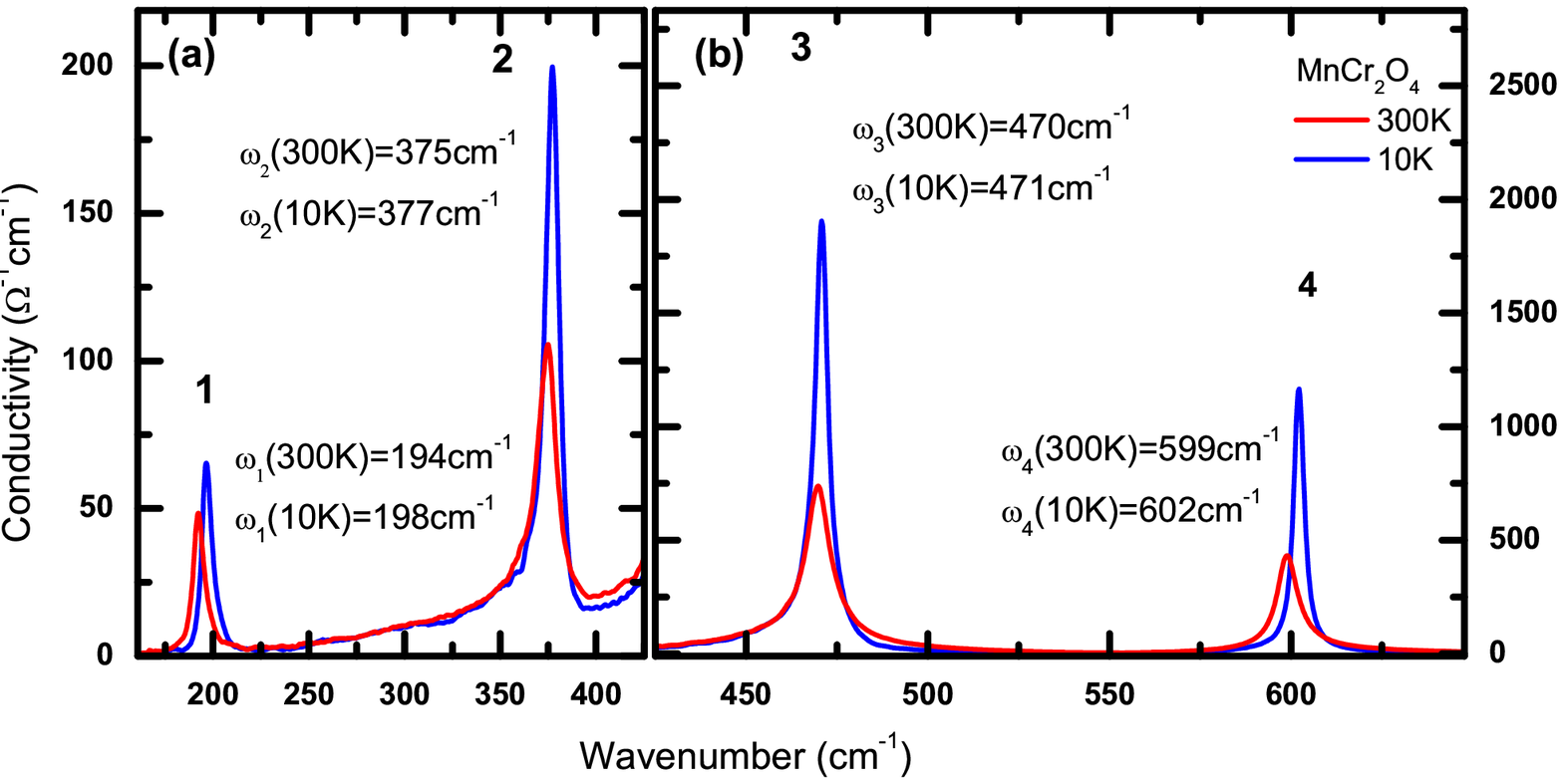}
    \caption{Optical conductivity spectra of MnCr$_2$O$_4$ at T=300\,K and 10\,K
            contain four distinct phonon modes. Neither the ferrimagnetic transition at T$_C$=51\,K nor the onset of conical spin order at T$_S$=14\,K are accompanied with an observable
            splitting of these $T_{1u}\mathrm{(IR)}$ phonon modes. Note that the conductivity scales for panel (a) and (b) differ by more than an order of magnitude.}
    \label{fig04}
    \end{figure}

    \begin{figure}[h!]
    \includegraphics[width=8.5truecm]{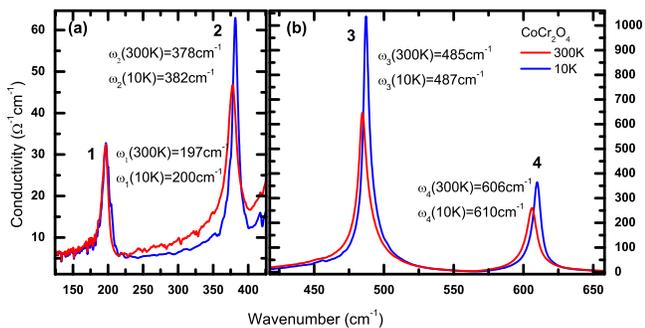}
    \caption{Optical conductivity spectra of CoCr$_2$O$_4$ at T=300\,K and 10\,K.
            Similarly to MnCr$_2$O$_4$, no splitting of the four $T_{1u}\mathrm{(IR)} $phonon modes can be traced either below the ferrimagnetic transition at T$_{C}$=93\,K
            or upon the conical spin ordering at T$_{S}$=26\,K.}
    \label{figA01}
    \end{figure}

The factor group analysis for cubic spinels (Fd$\overline{3}$m space
group, m3m (O$_h$) point group) yields 16 optical phonon modes in
the $\Gamma$ point:
    \begin{equation}
        \begin{array}{rcl}
         \Gamma(ACr_2O_4)_{Fd\overline{3}m}&=&A_{1g}\mathrm{(R)} \oplus 2A_{2u}\mathrm{(S)} \oplus E_g\mathrm{(R)} \oplus \\
                                          & &\oplus 2E_u\mathrm{(S)} \oplus T_{1g}\mathrm{(S)} \oplus 4T_{1u}\mathrm{(IR)} \oplus \\
                                          & &\oplus 3T_{2g}\mathrm{(R)} \oplus 2T_{2u}\mathrm{(S)},
    \end{array}
    \label{eq:cubic}
    \end{equation}
where abbreviations IR, R and S refer to infrared-active,
Raman-active and silent modes, respectively. Thus, in the cubic
phase four $T_{1u}\mathrm{(IR)}$ modes are expected to appear in the optical spectra, all of which have three-fold
degeneracy. When the lattice symmetry is reduced to tetragonal each of
these $T_{1u}$ lattice vibrations are split into an $A_{2u}\mathrm{(IR)}$ singlet and an
$E_{u}\mathrm{(IR)}$ doublet mode. Furthermore, the two originally silent
$T_{2u}$ phonon modes can split into an infrared-active $E_{u}\mathrm{(IR)}$
doublet and a silent $B_{2u}$ singlet. If the symmetry of the
lattice further lowered the degeneracy of each mode is completely lifted.

\begin{figure*}[t!]
    \includegraphics[width=11.3truecm]{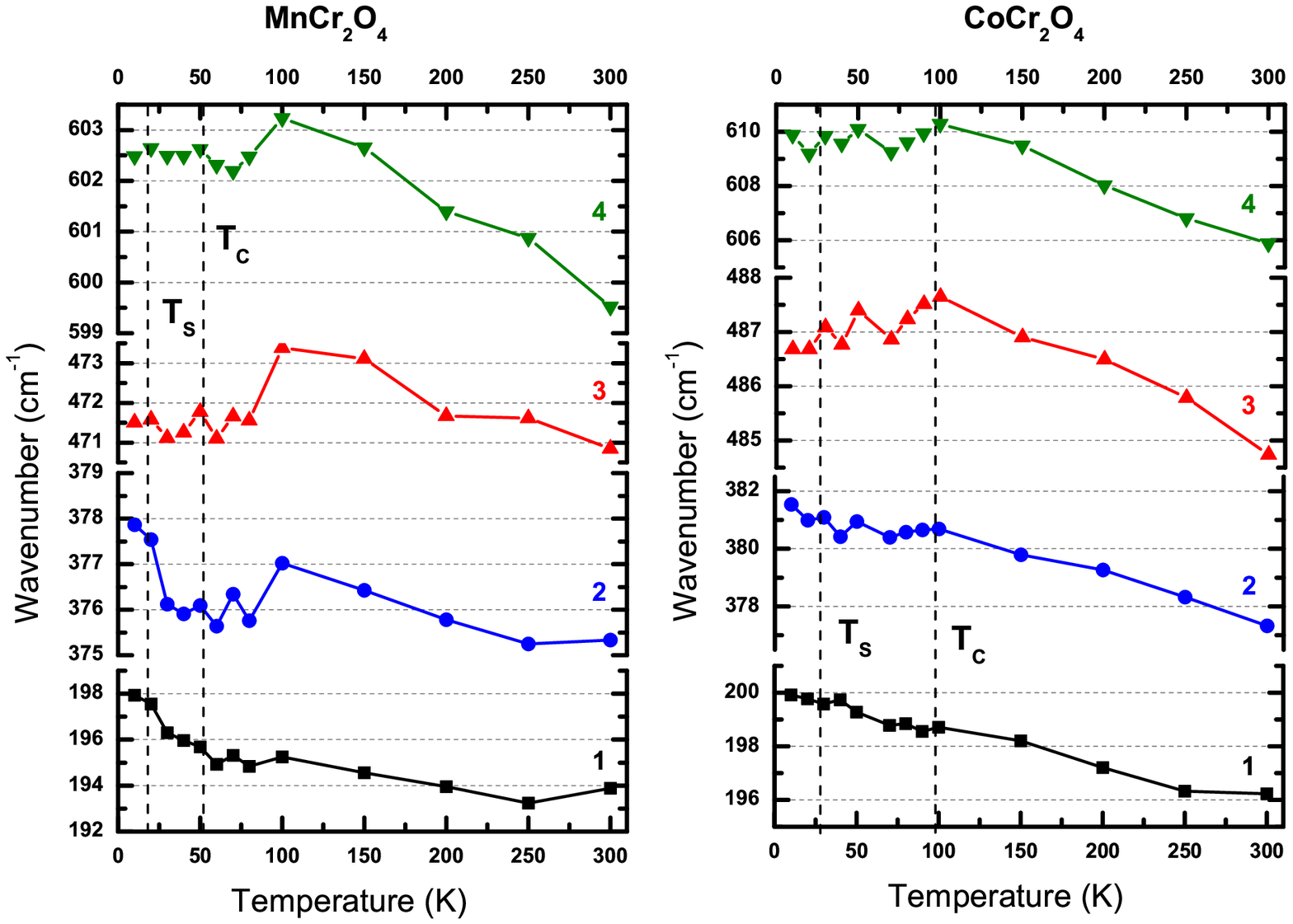}
    \caption{Temperature dependence of phonon mode energies in MnCr$_2$O$_4$ and CoCr$_2$O$_4$. Mn$^{2+}$ and Co$^{2+}$ ions, located on the A site of the spinel structure,
                have no orbital degeneracy (Fig.~\ref{fig01}(b)). Correspondingly, no structural transition associated with a cooperative Jahn-Teller distortion has been observed. No
                measurable splitting of the phonon modes occur upon the magnetic phase transitions either. T$_C$ and T$_S$ label the transition
                to a ferrimagnetic state and to the ground state with conical spin order, respectively.}
    \label{fig09}
    \end{figure*}

Optical conductivity spectra of MnCr$_2$O$_4$ and CoCr$_2$O$_4$,
spinels with no orbital degrees of freedom, are shown in
Fig.~\ref{fig04} and Fig.~\ref{figA01}, respectively. All four
infrared active $T_{1u}\mathrm{(IR)}$ optical phonon modes expected in the
cubic phase are observed at room temperature for both compounds. As
the temperature is decreased to T=10\,K the lifetime of each phonon
mode is increased by $\sim$30\,\%. The detailed temperature
dependence of the resonance frequencies was determined by fitting
the spectra according to Eq.~\ref{eq:oscmodel} and the corresponding
results are summarized in Fig.~\ref{fig09}.

All the modes show weak hardening towards lower temperatures
followed by a tiny softening associated with the ferrimagnetic
transition. Additionally, in MnCr$_2$O$_4$ the first and second mode
(located at $\sim$195\,cm$^{-1}$ and $\sim$376\,cm$^{-1}$,
respectively) are shifted toward higher energies upon the conical
spin ordering at $T_s$=14\,K. Although in CoCr$_2$O$_4$ an anomalous
broadening was observed for the third and fourth mode below
T$_C$=93\,K, no clear sign of phonon splitting can be discerned in
either of the two compounds within the resolution of our experiment,
$\sim 0.5\,cm^{-1}$. These results are compatible with a very weak,
if any, distortion of the cubic lattice in MnCr$_2$O$_4$ and
CoCr$_2$O$_4$ even at T=10\,K.

\subsection*{Phonon modes in ACr$_2$O$_4$ spinels with Jahn-Teller active ions}

\begin{figure}[h!]
    \includegraphics[width=8.5truecm]{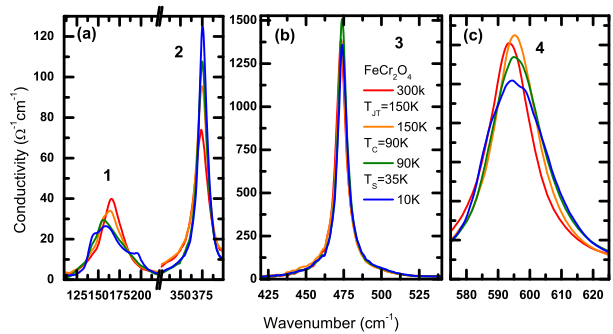}
    \caption{Optical conductivity spectra of FeCr$_2$O$_4$ at various temperatures in the vicinity of the Jahn-Teller transition at T$_{JT}$=135\,K,
            the ferrimagnetic ordering at T$_{C}$=93\,K and the onset of the conical order at T$_{S}$=35\,K.
            Lowering of the crystal symmetry can be traced as a temperature induced splitting of the lowest- and the highest-energy phonon modes both at T$_{JT}$ and T$_C$,
            while splitting of the other two $T_{1u}\mathrm{(IR)}$ modes cannot be resolved.}
    \label{figA02}
    \end{figure}

\begin{figure}[h!]
    \includegraphics[width=8.5truecm]{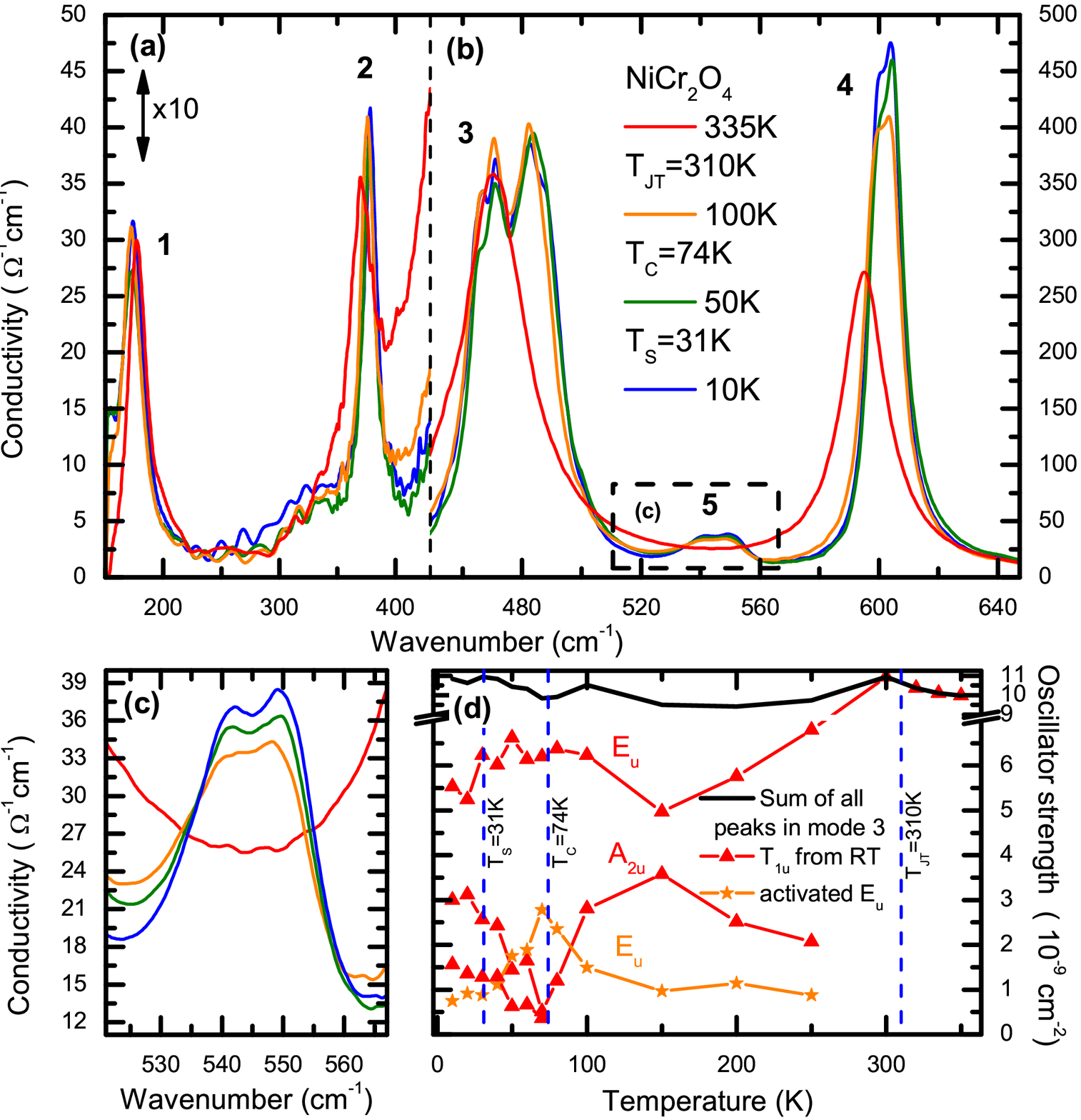}
    \caption{(a-b) Optical conductivity spectra of NiCr$_2$O$_2$ at various temperatures over the region of
            the Jahn-Teller transition at T$_{JT}=320$\,K, the ferrimagnetic ordering at T$_C=74$\,K and the onset of the canted magnetic structure
            at T$_S=31$\,K. Splitting of the third and the fourth $T_{1u}\mathrm{(IR)}$ modes can be observed both at
            T$_{JT}$ and T$_C$. Scale of low energy part is magnified ten times larger. (c) Between these modes the activation of an $E_u\mathrm{(IR)}$ mode, which originates from a
            $T_{2u}\mathrm{(S)}$ mode being silent in the cubic phase, can be followed. Rigorous analysis shows the activation of another
            $E_u\mathrm{(IR)}$ mode upon $T_{JT}$, which is located among the branches of the third $T_{1u}\mathrm{(IR)}$ mode.
            (d) Oscillator strength for this $E_u\mathrm{(IR)}$ mode and for the branches of the third $T_{1u}\mathrm{(IR)}$ mode
            as a function of temperature.}
    \label{fig07}
    \end{figure}

In FeCr$_2$O$_4$ and NiCr$_2$O$_4$ spinel oxides with Jahn-Teller
active A$^{2+}$ ions, the presence of four $T_{1u}\mathrm{(IR)}$ phonon modes
above room temperature, respectively followed in Fig.~\ref{figA02}
and Fig.~\ref{fig07}, indicates the stability of the high-symmetry
cubic spinel structure. However, upon the Jahn-Teller transition of
FeCr$_2$O$_4$ at T$_{JT}$=135\,K the lowest- and the highest-energy
mode, while in NiCr$_2$O$_4$ below T$_{JT}$=320\,K the two
high-energy modes are split indicating structural transition to a tetragonal state.

Besides the splitting of the third and fourth mode in NiCr$_2$O$_4$,
a new excitation appears below the Jahn-Teller transition at
$\omega_5$=545\,cm$^{-1}$ (see Fig.~\ref{fig07}(c)). This fifth mode
has low oscillator strength, short lifetime and cannot be deduced
from either of the cubic $T_{1u}\mathrm{(IR)}$ phonons. However, in the
tetragonal phase the originally silent $T_{2u}\mathrm{(S)}$ phonon mode can
split into an infrared-active $E_{u}\mathrm{(IR)}$ doublet and a silent
$B_{2u}\mathrm{(S)}$ singlet beside the splitting of the $T_{1u}\mathrm{(IR)}$ modes
to $A_{2u}\mathrm{(IR)}$ $\oplus$ $E_{u}\mathrm{(IR)}$. We assign this weak mode
around $\omega_5$=545\,cm$^{-1}$ to such an $E_{u}\mathrm{(IR)}$ mode
originating from one of the two $T_{2u}\mathrm{(S)}$ modes silent in the
cubic symmetry. Another mode with similarly small intensity emerges
just below T$_{JT}$ in the vicinity of the third $T_{1u}\mathrm{(IR)}$ mode.
We assign it as the other $E_{u}\mathrm{(IR)}$ mode activated by the
tetragonal splitting. The fact that the third $T_{1u}\mathrm{(IR)}$ is split
into two branches at T$_{JT}$ and becomes three non-degenerate modes
with comparable oscillator strength below T$_C$ highly support this
assignment, i.e. it shows the independent origin of this weak mode.

    \begin{figure}[h!]
    \includegraphics[width=8.5truecm]{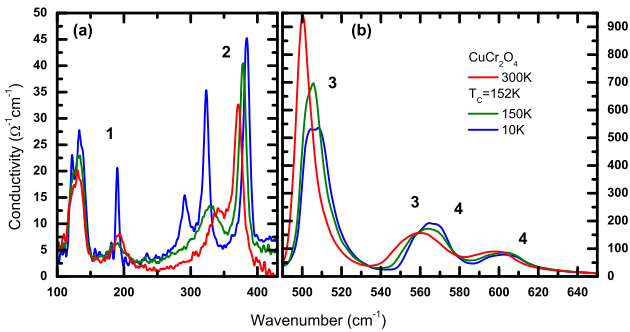}
    \caption{Optical conductivity spectra of CuCr$_2$O$_4$ at various temperatures. At room temperature all T$_{1u}\mathrm{(IR)}$ phonon modes are split
            due to the Jahn-Teller transition at T$_{JT}=$854\,K. For each mode the remaining degeneracy is lifted upon the magnetic ordering at T$_C=152$\,K.}
    \label{figA03}
    \end{figure}

    \begin{figure*}[t!]
    \includegraphics[width=15.7truecm]{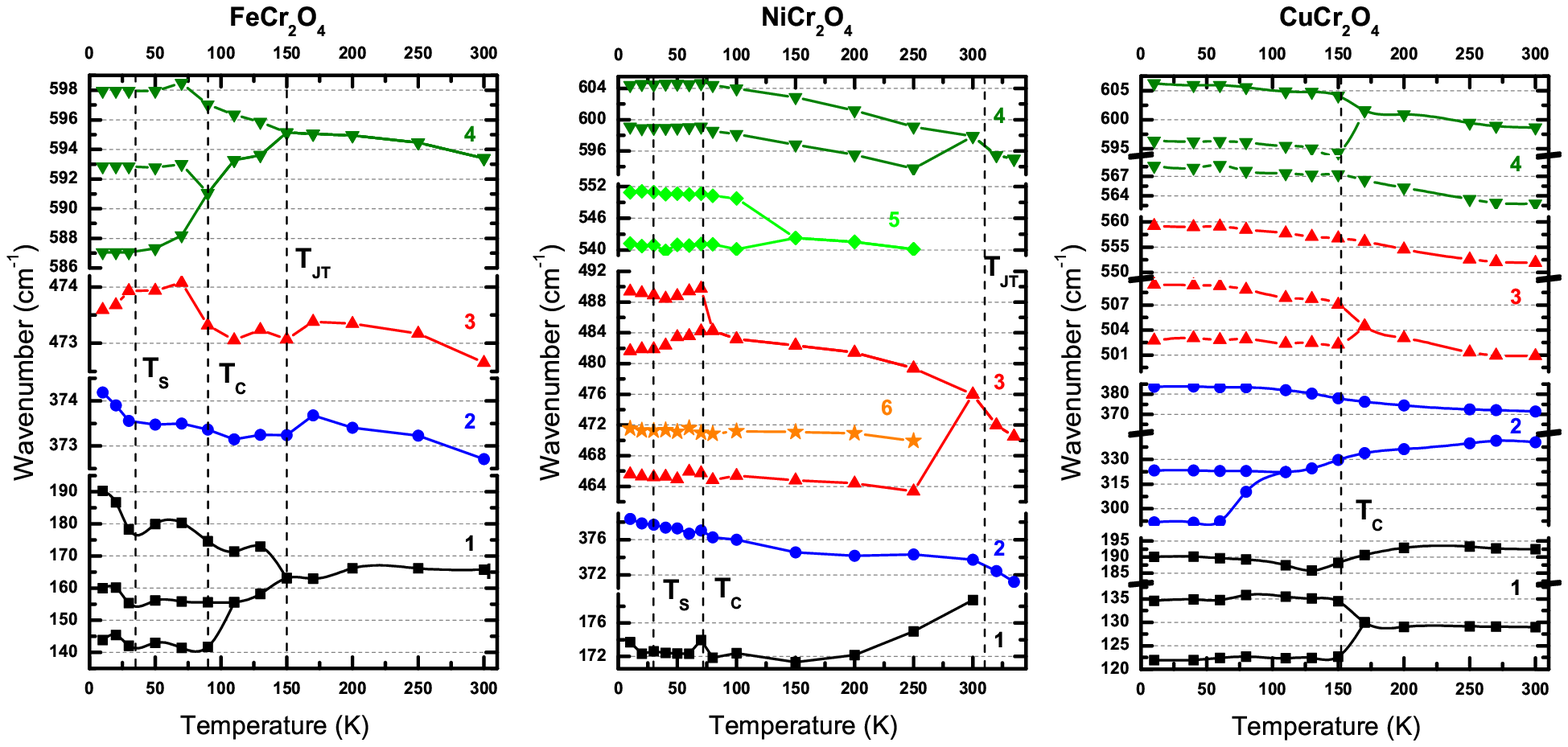}
    \caption{Temperature dependence of phonon energies in FeCr$_2$O$_4$, NiCr$_2$O$_4$ and CuCr$_2$O$_4$. Lattice vibrations
                originating from the same $T_{1u}$ and $T_{2u}$ cubic modes are labeled with the same color and symbol.
                The temperature of the phase transitions are indicated by dashed lines.}
    \label{fig08}
    \end{figure*}

The optical conductivity spectra of CuCr$_2$O$_4$ are shown in
Fig.~\ref{figA03}. Each $T_{1u}\mathrm{(IR)}$ mode is already split to
$A_{2u}\mathrm{(IR)}$ $\oplus$ $E_{u}\mathrm{(IR)}$ at room temperature in accordance
with its cubic to tetragonal phase transition at T$_{JT}$=854\,K.
The remaining twofold degeneracy of each $E_{u}\mathrm{(IR)}$ is lifted upon
the ferrimagnetic transition. In contrast to NiCr$_2$O$_4$, no
activation of silent modes can be followed.

A summary about the temperature dependence of the phonon frequencies
are plotted in Fig.~\ref{fig08} for the spinel oxides with orbital
degeneracy, A=Fe, Ni and Cu. The ferrimagnetic phase transition
completely removes the threefold degeneracy of some of the cubic
$T_{1u}\mathrm{(IR)}$ modes in all the three compounds, therefore, the cubic
symmetry of the lattice is lower than tetragonal in the ground state
of these compounds. The modes whose degeneracy is partly lifted by
the Jahn-Teller distortion exhibit further splitting at the
subsequent magnetic transition at T$_C$ except for the
highest-energy mode in NiCr$_2$O$_4$. The magnetically induced
splitting is as large as $\Delta\omega$ / $\omega$ $\approx$10$\%$
for the lower-energy vibrations. The sensitivity of the phonon modes
to the magnetic ordering depends on the chemical composition; in
FeCr$_2$O$_4$ the first and the fourth modes, in NiCr$_2$O$_4$ the
third mode, while in CuCr$_2$O$_4$ all modes are split. Furthermore,
the ferroelectric phase transition in FeCr$_2$O$_4$ is accompanied
by a sudden hardening of the lower-energy modes. The partial
activation of two silent $T_{2u}\mathrm{(S)}$ vibrations was only observed in
NiCr$_2$O$_4$. Among these new $E_u\mathrm{(IR)}$ modes the vibration around
$\omega_5$=545\,cm$^{-1}$ is sensitive to the magnetic ordering as
it shows a splitting in the vicinity of T$_C$. As already reported
in the literature \cite{Siratori}, the lowest-energy mode in this
compound exhibit an anomalous softening with decreasing temperature.
Such unusual temperature dependence of the low-energy vibrations was
also found in FeV$_2$O$_4$ \cite{Siratori}.

\subsection*{Intra-3d-band excitations of A site ions}

Weak in-gap excitations observed over the mid-infrared--visible range in ACr$_2$O$_4$ spinels are interpreted as transitions within the \textit{3d} levels of the A-site cations split by the local crystal field, the Coulomb interaction and the spin-orbit coupling \cite{Ohgushi2008}. In ACr$_2$O$_4$ compounds these parity-forbidden transitions become allowed due to the lack of inversion symmetry in the local tetrahedral environment which provides the possibility to investigate the electronic structure of the A site ions by optical spectroscopy. In the subsequent section the optical conductivity spectra are analysed in terms of ligand field theoretical calculations.

We observed \textit{3d}-\textit{3d} excitations in the optical conductivity spectra of A=Cu, Ni, Co and A=Fe chromium spinel oxides, while no such transition was found in MnCr$_2$O$_4$ (see Fig.~\ref{fig02}). The absence of this excitation in the latter compound indicates that high-spin state is realized for the A-site since there is no spin-allowed \textit{3d}-\textit{3d} transition within the half-filled d-band of the Mn$^{2+}$ ion in the presence of strong Hund-coupling as shown in Fig.~\ref{fig01}(b).

\begin{figure*}[t!]
    \includegraphics[width=15.7truecm]{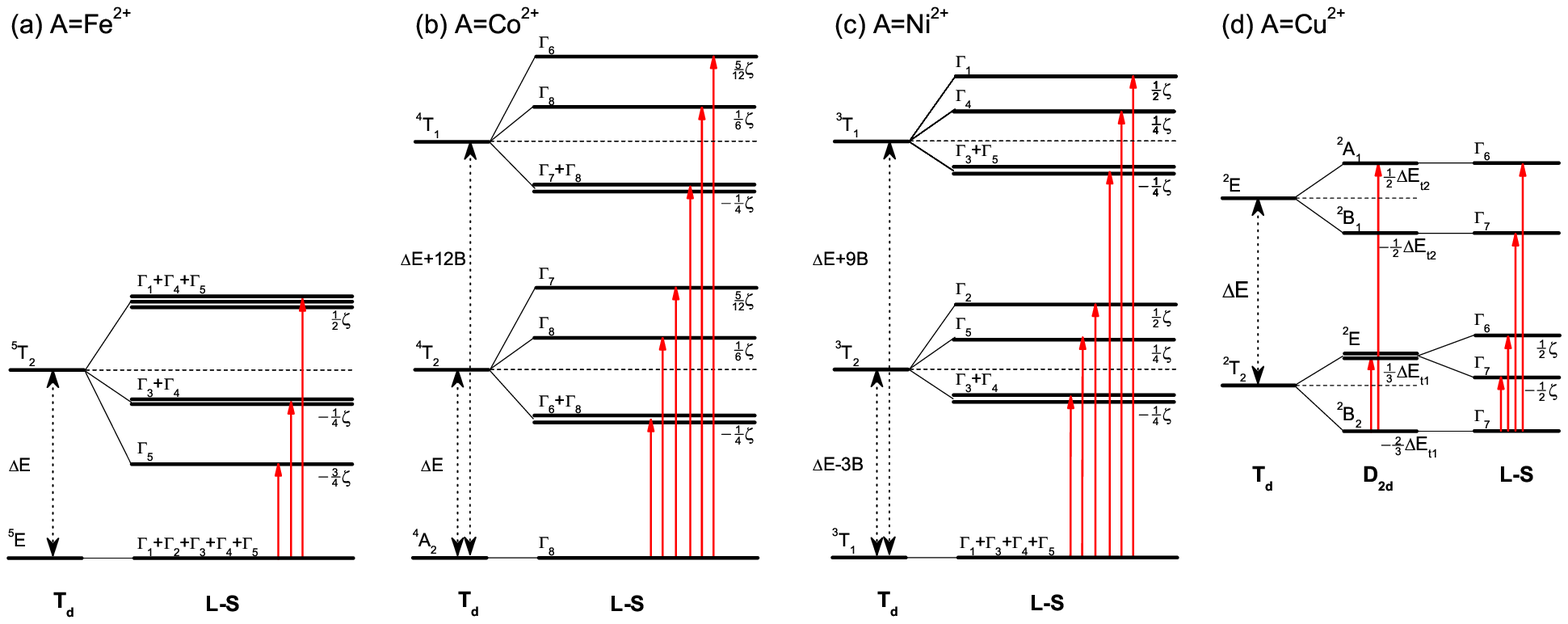}
    \caption{Energy level diagrams corresponding to Fe$^{2+}$, Co$^{2+}, $Ni$^{2+}$ and Cu$^{2+}$ 3d orbitals occupied by holes. In these diagrams only the spin-allowed single-particle excitations are shown, red arrows represents the allowed optical transitions. (a-b) \textit{3d}-\textit{3d} optical transitions within the d-band of the A-site ion split by the cubic (T$_d$) crystal field, $\Delta$E, the Racah parameter, B and the spin-orbit interaction, $\zeta$ for A=Fe$^{2+}$ and  A=Co$^{2+}$ \cite{Ohgushi2008,Sugano}. (c) In case of A=Ni$^{2+}$ the effect of local T$_d$ symmetry and spin-orbit interaction was taken into account, the latter as the weakest perturbation. Here, symmetry lowering associated with the weak tetragonal splitting was neglected. (d) In case of A=Cu$^{2+}$ the effect of Jahn-Teller distortion, i.e. tetragonal splitting, $\Delta$E$_1$ was considered stronger than the spin-orbit interaction.}
    \label{fig3d3d01}
    \end{figure*}

Ohgushi et al.~have discussed the case of A=Fe and Co in the frame of ligand field theory including spin-orbit coupling \cite{Ohgushi2008}. The main results of their calculations are reproduced in the first two panel of Fig.~\ref{fig3d3d01}. The transition in FeCr$_2$O$_4$ at around $\sim$0.49\,eV, was assigned as a single electron transition through the crystal field gap, $^5$E(e$^3$t$_2^3$) $\rightarrow$ $^5$T$_2$(e$^2$t$_2^4$), while the two absorbtion peaks in CoCr$_2$O$_4$ at around $\sim$0.84\,eV and $\sim$1.95\,eV correspond to the
$^4$A$_2$(e$^4$t$_2^3$) $\rightarrow$ $^4$T$_2$(e$^3$t$_2^4$) and
$^4$A$_2$(e$^4$t$_2^3$) $\rightarrow$ $^5$T$_1$(e$^3$t$_2^4$)
transitions, respectively. From these excitations the crystal field splitting corresponding to the cubic state and the Racah parameter, B, characteristic to the Coulomb interaction can be determined for Co$^{2+}$: $\Delta$E=0.84\,eV and B=0.083\,eV, respectively.

\begin{figure}[h!]
    \includegraphics[width=8.5truecm]{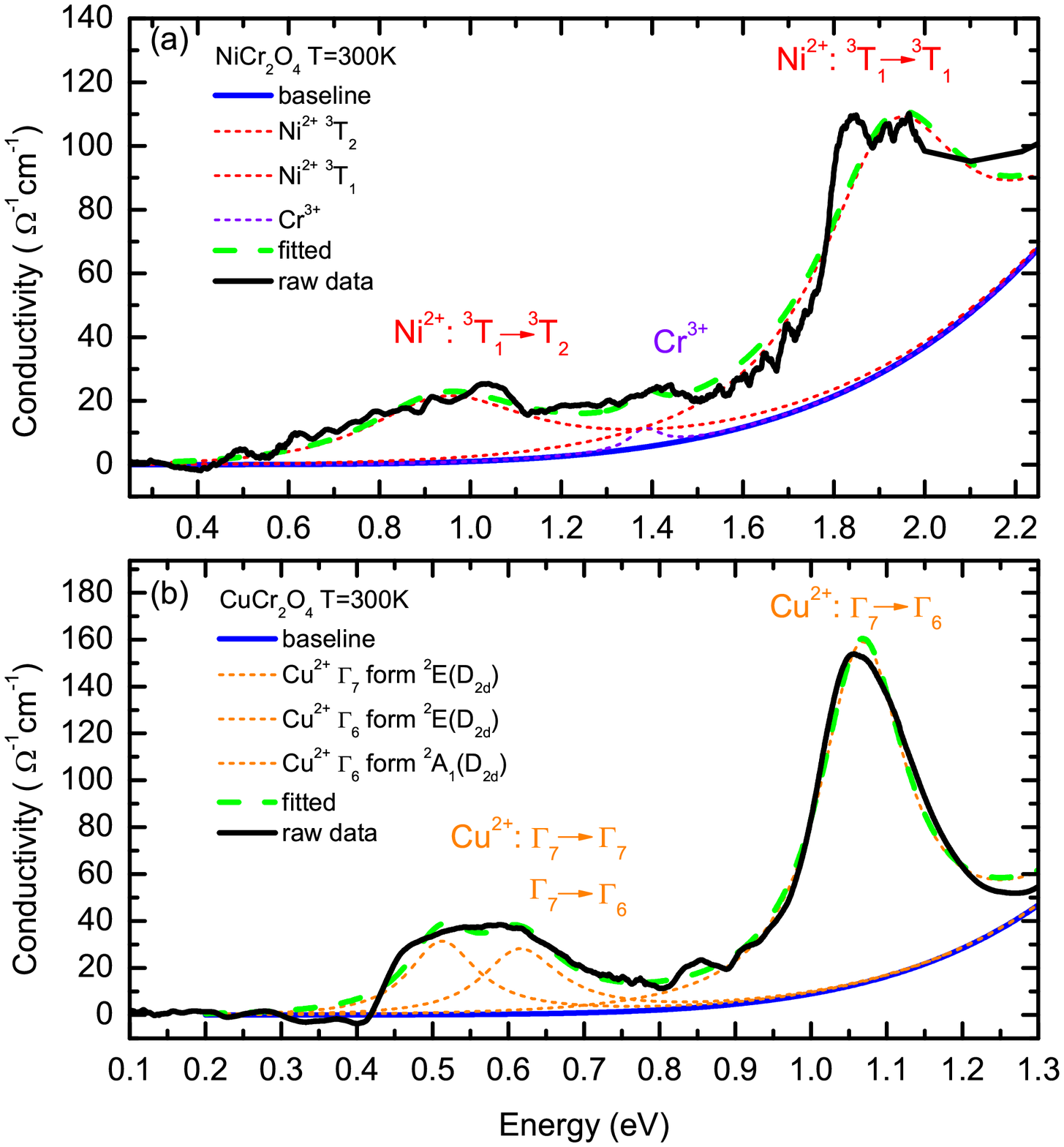}
    \caption{Analysis of the intra-atomic \textit{3d}-\textit{3d} transitions in NiCr$_2$O$_4$ and CuCr$_2$O$_4$ as shown in panel (a) and (b), respectively. Labels correspond to notation introduced in Fig.~\ref{fig3d3d01}. Besides the oscillators, there is a common baseline corresponding to the tail of the band edge. }
    \label{fig12}
    \end{figure}

In the low-temperature optical conductivity spectrum of NiCr$_2$O$_4$ two distinct structures are visible around $\sim$0.95\,eV and $\sim$1.94\,eV, which can be assigned as \textit{3d}-\textit{3d} transitions of the tetrahedrally coordinated Ni$^{2+}$. Since the fine structure of these peaks cannot be resolved, cubic (T$_d$) site symmetry was assumed in our model and the tetragonal distortion was neglected \cite{Sugano}. The energy levels of the eight 3d electron of the Ni$^{2+}$, relevant for the single particle excitations, is depicted in Fig.~\ref{fig3d3d01}(c). The two peaks in the optical conductivity spectrum were fitted with two Lorentzian corresponding to the two electric-dipole-allowed transitions: $^3$T$_1$(e$^4$t$_2^4$) $\rightarrow$ $^3$T$_2$(e$^3$t$_2^5$) and
$^3$T$_1$(e$^4$t$_2^4$) $\rightarrow$ $^3$T$_1$(e$^3$t$_2^5$) (see Fig.~\ref{fig12}(a)). The cubic crystal field splitting and the Racah parameter for the Ni$^{2+}$ ion were deduced from the transition energies: $\Delta$E=1.20\,eV and B=0.082\,eV, respectively. The spin-orbit and the unresolved tetragonal splitting may be responsible for the large width of the resonances peaks.

In contrast to NiCr$_2$O$_4$, the crystal lattice of the CuCr$_2$O$_4$ has strong tetragonal distortion at low temperature \cite{Crottaz1997}. Perhaps, the low-symmetry of the lattice is responsible for the large number of peaks observed in the near infrared and visible part of the optical conductivity spectrum. As in other ACr$_2$O$_4$ compounds the excitations around E=1.6\,eV are assigned to the intra-atomic \textit{3d}-\textit{3d} transitions of Cr$^{3+}$ split by the tetragonal distortion. Compared to the sister compounds the strength of this transition is remarkably enhanced. The two lower-energy peaks at around $\sim$0.56\,eV and $\sim$1.06\,eV are related to the \textit{3d}-\textit{3d} transitions of Cu$^{2+}$. Due to the strong Jahn-Teller distortion in the ligand field calculations tetragonal D$_{2d}$ local symmetry was assumed and the spin-orbit coupling was also taken into account. The energy levels and the possible optical transitions are shown in Fig.~\ref{fig3d3d01}(d). In the D$_{2d}$ symmetry the $^2$B$_2$ $\rightarrow$ $^2$B$_1$ transition is forbidden, and only the spin-orbit interaction makes it weakly allowed, therefore, its oscillator strength was assumed to be zero. The optical conductivity spectrum was fitted with three Lorentzian oscillator as shown in Fig.~\ref{fig12}(b). The obtained values for the cubic and the tetragonal crystal field splitting, and the spin-orbit coupling are $\Delta$E\,=\,0.69\,eV, $\Delta$E$_{t1}$\,=\,0.56\,eV and $\zeta$\,=\,0.1\,eV, respectively, if the tetragonal splitting within the cubic $^2$E term, $\Delta$E$_{t2}$ is neglected.

The different parameters of the electronic structure of the A-site ions, namely the cubic and tetragonal crystal field splittings, $\Delta$E and $\Delta$E$_t$, respectively, the Racah parameter, B and the strength of the spin-orbit coupling, $\zeta$ are summarized in Table \ref{table:CFparameters} for the investigated ACr$_2$O$_4$ spinel oxides. These fundamental energy scales may serve as a starting point for a microscopic theory describing the magneto-elasticity of the ACr$_2$O$_4$ spinels and provide in general an important input to theories describing their magnetic states.

\begin{table}[h]
\caption{\textit{Cubic ($\Delta$E) and tetragonal ($\Delta$E$_t$) crystal field splittings, the Racah parameter (B) and the strength of the spin-orbit coupling ($\zeta$) as determined in the present study for various spinel oxides with magnetic A-site ions.}}
\label{table:CFparameters}
\begin{center}
\begin{tabular}{|c|c|c|c|c|}
\hline
  & Fe$^{2+}$\cite{Ohgushi2008} & Co$^{2+}$\cite{Ohgushi2008} & Ni$^{2+}$ & Cu$^{2+}$ \\
\hline
$\Delta$E (eV) & 0.49 & 0.84 & 1.20 & 0.69 \\
\hline
$\Delta$E$_{t1}$, $\Delta$E$_{t2}$ (eV)  &   &   &   & 0.56 , $\sim$0 \\
\hline
B (eV) &   & 0.083 & 0.082 &   \\
\hline
$\zeta$ (eV) & 0.13 ($^5$T$_2$) & 0.25 ($^4$T$_2$) & & 0.1 ($^2$T$_2$) \\
	&	& 0.34 ($^4$T$_1$) &  & \\
\hline
\end{tabular}
\end{center}
\end{table}

\section*{Discussion}

From the viewpoint of the dynamical properties of the lattice,
ACr$_2$O$_4$ chromium spinels (A=Mn, Fe, Co, Ni and Cu) can be
classified into two distinct groups. The main results of our optical
study about these classes are summarized in Fig.~\ref{fig09} and
Fig.~\ref{fig08}.

In the first group the A-site ions have no orbital degeneracy (A=Mn
and Co) and neither of these compounds show observable phonon mode
splitting upon the magnetic phase transitions from room temperature
down to T$=$10\,K. Anomalous broadening of the third and fourth mode
of CoCr$_2$O$_4$ below T$_C$=93\,K is observed, which may indicate a
mode splitting slightly below the spectral resolution of our
experiment ($\sim$0.5\,cm$^{-1}$). In MnCr$_2$O$_4$ we could not
even follow any broadening of the modes associated with the magnetic
transitions. Therefore, we conclude that the magnetoelasticity is
weak in these compounds and the cubic symmetry of the lattice is
nearly preserved even in their magnetically ordered ground state.

For materials in the second class -- where A-site ions are
Jahn-Teller active, i.e. they have orbital degeneracy in the cubic
phase -- we observed phonon-mode splitting associated with the
cooperative Jahn-Teller distortion. The sequence of the splittings
shows that a cubic to tetragonal distortion takes place at T$_{JT}$.
Phonon modes in CuCr$_2$O$_4$ exhibit a relative splitting of
$\approx$30\%, $\approx$8\%, $\approx$9\% and $\approx$7\% in
ascending order of phonon energies already at room temperture
\cite{Bordacs_article}. For comparison, splitting of the third and
fourth phonon modes in NiCr$_2$O$_4$ is $\approx$4\% and
$\approx$1\%, while the lowest- and highest-energy phonons in
FeCr$_2$O$_4$ are split by $\Delta \omega / \omega_o \approx$20\%
and $\approx$1\%, respectively. Most of the modes whose degeneracy
is partly lifted by the cooperative Jahn-Teller distortion or those
activated upon the splitting of silent $T_{2u}$ lattice vibrations,
as is the case in NiCr$_2$O$_4$, are further split upon the
ferrimagnetic ordering. The magnitude of the magnetically induced
splitting is in the range of $\approx$10\% for low-energy and
$\approx$1\% for high-energy phonon modes, the large splitting of
low-energy modes implies strong spin-lattice coupling for these
lattice vibrations resulting in the strong magnetoelastic effects. Recently, the static orthorhombic distortion of the crystalline lattice upon the magnetic phase transition has also been observed by means of high-resolution X-ray diffraction in NiCr$_2$O$_4$ and CuCr$_2$O$_4$ \cite{Ishibashi2006,Suchomel2012}.

In conclusions, we found that in ACr$_2$O$_4$ spinel compounds --
where the frustration of the pyrochlore sublattice is lifted by
long-range magnetic ordering due to the exchange interaction between
the spins of the A$^{2+}$ and Cr$^{3+}$ ions -- the criterion of
strong magnetoelasticity is the orbital degeneracy of the A-site ion
in the high-symmetry cubic phase. This observation is valid
irrespective to the details of the magnetic order, i.e. we found
strong magnetoelasticity for the compounds FeCr$_2$O$_4$,
NiCr$_2$O$_4$ and CuCr$_2$O$_4$ independently whether they have
conical or canted spin order in their ground state. A theoretical model reproducing the effect of orbital occupancy on the magnetoelasticity of these compounds will be published elsewhere \cite{tobepublished}. Moreover, based on the
analysis of the intraatomic \textit{3d}-\textit{3d} transition of the A-site ions, we
could determine the following fundamental energies: crystal-field
splitting in the cubic and tetragonal phases, the Racah parameters
representing the strength of the intraatomic Coulomb repulsion and
the spin-orbit interaction in some cases.

\section*{Acknowledgement}
This work was supported by Hungarian Research Funds OTKA PD75615, Bolyai 00256/08/11, T\'AMOP-4.2.2.B-10/1-2010-0009 and it was partly supported by Grant-in-Aid for Scientific Research (S) No. 24224009, and FIRST Program by the Japan Society for the
Promotion of Science (JSPS). Financial support by the Deutsche Forschungsgemeinschaft through SFB 484 is acknowledged.


\begin{thebibliography}{99}
\bibitem{Ohgushi_talk}
    High energy spectra of MnCr$_2$O$_4$ was used to extend the reflectivity of NiCr$_2$O$_4$ above 3eV.
\end{thebibliography}
\end{document}